\documentclass[12pt]{iopart}
\expandafter\let\csname equation*\endcsname\relax
\expandafter\let\csname endequation*\endcsname\relax
\usepackage{amssymb}
\usepackage{amsmath}
\usepackage{amsthm}
\usepackage[numbers,sort&compress,square]{natbib}
\usepackage{bm}
\usepackage{graphicx}
\usepackage{calc}
\usepackage[dvipsnames]{xcolor}
\usepackage{hyperref}
\usepackage[capitalise]{cleveref}
\usepackage{algcompatible}

\newcommand{\mean}[1]{\left\langle#1\right\rangle}

\begin{document}

\title[Tolerance in Kolkata Paise Dining Club]{Multi-Group Dynamics with Tolerant Switching in the Kolkata Paise Restaurant Problem with Dining Clubs}

\author{Akshat Harlalka$^1$ and Christopher Griffin$^2$}
\address{
	$^1$Department of Computer Science,
    The Pennsylvania State University, 
    University Park, PA 16802}
\address{
	$^2$Applied Research Laboratory,
	The Pennsylvania State University,
    University Park, PA 16802
    }

\eads{\mailto{avh5936@psu.edu}, \mailto{griffinch@psu.edu}}
\date{\today~-~Preprint}
\begin{abstract}
We study the Kolkata Paise Restaurant Problem (KPRP) with multiple dining clubs, extending work in [A. Harlalka, A. Belmonte and C. Griffin, \textit{Physica A}, 620:128767, 2023]. In classical KPRP, $N$ agents chose among $N$ restaurants at random. If multiple users choose the same restaurant, only one will eat. In a dining club, agents coordinate to avoid choosing the same restaurant, but may collide with users outside the club. We consider a dynamic in which agents switch among clubs or the unaffiliated (free agent) group based on their comparative probability of eating. Agents' affiliations are sticky in the sense that they are insensitive (tolerate) to differences in eating probability below a threshold $\tau$ without switching groups. We study the tendency of one group (dining club or free agent group) to become dominant as a function of tolerance by studying the mean-field dynamics of group proportion. We then show empirically that the mean-field group dynamic (assuming infinite populations) differs from the finite population group dynamic. We derive a mathematical approximation in the latter case, showing good agreement with the data. The paper concludes by studying the impact of (food) taxation, redistribution and freeloading in the finite population case. We show that a group that redistributes food tends to become dominant more often as a function of increasing tolerance to a point, at which point agents do not switch frequently enough to enable group dynamics to emerge. This is negatively affected by freeloaders from the non-redistributing group. 
\end{abstract}

\vspace{2pc}

\maketitle

\section{Introduction}

Chakrabarti et al. \cite{CMC07} first introduced the Kolkata Paise Restaurant Problem (KPRP) in 2007. In this problem, one assumes that $N \gg 1$ agents choose from among $N$ restaurants. There may be an implicit ranking of the restaurants, or the restaurants may be considered equivalent. The ranking determines a payoff for eating at a given restaurant. If two or more agents choose the same restaurant, then one is served at random by the restaurant. This is a simple model of both the paise food vendors common in Kolkata and also describes an abstract decentralized resource allocation problem. Variations of this problem have been studied extensively in the econophysics literature \cite{HBG2023, BGCN12,CCGM17,CG19,GC17,DSC11,BMM13,BM21,SC20,GDCM12,CCCM15,CRS22,KPA22,CMC07,MK17,R13,Y10,GCCC14}. 

KPRP is an anti-coordination game (like Chicken or Hawk-Dove) \cite{M12}. Minority games are also examples of this class of games \cite{CZ98,HZDH12}, as is the El Farol bar problem \cite{arthur1994, decara1999, FGH02, challet2004}. These types of games  emerge as models of channel sharing in communications systems \cite{AFGJ13,AFGJ13a,GK14}. A complete survey of the KPRP can be found in \cite{CCGM17,BMM13,CCCM15}. When all restaurants are ranked equally (i.e., have payoff $1$) and agents choose a restaurant at random, the expected payoff to each agent is known to approach $1-1/e$ as $N \to \infty$. Using randomized strategies and resource usage analysis, the mean payoff can be increased to $ \sim 0.8$ \cite{GCMC10}. Identifying resource sharing mechanisms to improve on the uncoordinated outcome is a central problem in KPRP.

Distributed and coordinated solutions to optimize agent payoff are discussed in \cite{CG19,GC17,DSC11,KPA22}. Several authors have considered classical learning in the KPRP \cite{BSC24,CRS22,GCMC10,GSC10} while quantum learning is considered in \cite{CRS22}. Quantum variants of the problem are also considered in \cite{CRS22,R13,Y10}, with applications to other areas of physics presented in \cite{BM21,GCCC14,MK17,GDCM12}. Phase transitions in the KPRP have been studied in \cite{BGCN12,SC20}. 

In \cite{HBG2023}, Harlalka et al. introduced the idea of a coordinated dining club into the KPRP. In this construct, a (small) set of agents agrees to collude to avoid restaurant collision. If agents are given the choice to join or leave the dining club, we showed that universal acceptance of the centrally coordinated dining club is globally evolutionarily stable even when the club members redistribute food (resources) to those members who do not eat because of collision with a non-member. This stability becomes local when non-members are allowed to freeload on redistributed food. 

In this paper, we consider a scenario in which multiple dining clubs have formed and agents are free to choose among them. Choice is governed by an estimate of the proportion of time an agent eats when compared to another random agent from an alternate club. To model more realistic decision-making, we assume agents have a ``threshold for inequality tolerance'' where even if another agent eats with higher estimated probability, they will not switch clubs because of friction (e.g., intangible payoffs, loyalty etc.). We then study the impact this tolerance to inequality has on the stability of the dining clubs. In particular, we study differences that arise between the continuum limit, when the number of agents goes to infinity, when compared to observed (simulated) behaviours in the finite population case. We obtain mathematical models for both scenarios. An interesting finding is that in the absence of freeloading, dining clubs that redistribute food obtain a slight advantage when tolerance to inequality is between two thresholds in the finite population case.

The remainder of this paper is organized as follows: In \cref{sec:MathModel} we provide a mathematical model for our problem setup and study the continuum limit case when the number of agents goes to infinity. We introduce the simulation functions used for experimentation in \cref{sec:SimFunctions}. Dining club evolution in finite populations without redistribution is studied in \cref{sec:NoRedist}. We then study the impact of resource redistribution and freeloading in \cref{sec:Redistribution}. Conclusions and future directions are studied in \cref{sec:Conclusions}.

\section{Mathematical Model of Infinite Populations} \label{sec:MathModel}
We begin by analysing the KPRP with two dining clubs and a group of free agents. The analysis presented here could easily be extended to $N > 2$ dining clubs, however, as $N \to \infty$, we revert to the original KPRP with dining clubs replacing individual agents. For this work, we strictly consider the problem of two competing dining clubs and a group of free agents and study the evolution of the various groups through time. 

Let $n$ be the total size of the free group and $g_i$ ($i \in \{1,2\})$ be the sizes of the two dining clubs respectively. A member of dining club 1 eats if and only if they choose a restaurant not chosen by a member of dining club 2 or the free group \textit{or} if this occurs, they are randomly selected to eat from the group of individuals that arrive at the restaurant in question. Using this information, we can derive an expression for the probability that a random member of group 1 eats in any given round as,
\begin{equation}
p(n, g_1, g_2) = \sum_{k=0}^n\left[
\binom{n}{k}\left(\frac{1}{n+g_1+g_2}\right)^k
\left(\frac{n+g_1+g_2 - 1}{n+g_1+g_2} \right)^{n-k} P(n, g_1, g_2)
\right],
\label{eqn:p}
\end{equation}
where 
\begin{equation*}
P(n, g_1, g_2) = 
\left(\frac{n+g_1}{n+g_1+g_2}\right)\frac{1}{k+1} + 
\left(\frac{g_2}{n+g_1+g_2}\right)\frac{1}{k+2}.
\end{equation*}
Symmetrically, the probability that a random member from group 2 eats in any given round is, $p(n,g_2,g_1)$. In \cref{eqn:p}, 
\begin{equation*}
\binom{n}{k}\left(\frac{1}{n+g_1+g_2}\right)^k
\left(\frac{n+g_1+g_2 - 1}{n+g_1+g_2} \right)^{n-k},
\end{equation*}
gives the probability that $k$ members of the free group simultaneously select the same restaurant as a member of group 1. The expression,
\begin{equation*}
\left(\frac{n+g_1}{n+g_1+g_2}\right)\frac{1}{k+1} = 
\left(1 - \frac{g_2}{n+g_1+g_2}\right)\frac{1}{k+1},
\end{equation*}
gives the probability of the member of group 1 not colliding with a member of group 2 and being randomly selected to eat among the $k+1$ total visitors to the chosen restaurant. Lastly, 
\begin{equation*}
\left(\frac{g_2}{n+g_1+g_2}\right)\frac{1}{k+2},
\end{equation*}
gives the probability that the member of group 1, who collides with a member of group 2, is randomly selected to eat among the $k+2$ total visitors to the chosen restaurant. As noted, the computation (and hence explanation) is symmetric for a randomly chosen member of group 2.

The probability that a free group member eats can be computed as,
\begin{equation}
q(n,g_1,g_2) = \sum_{k=0}^{n-1}\left[
\binom{n-1}{k}\left(\frac{1}{n+g_1+g_2}\right)^k
\left(\frac{n+g_1+g_2 - 1}{n+g_1+g_2} \right)^{n-k-1} Q(n, g_1, g_2)
\right],
\label{eqn:q}
\end{equation}
where,
\begin{multline*}
Q(n, g_1, g_2) = \left[
\frac{n+g_1}{n+g_1+g_2}\frac{n+g_2}{n+g_1+g_2}\frac{1}{k+1} +\right.\\ \left.\left(
\frac{g_1}{n+g_1+g_2}\frac{n+g_1}{n+g_1+g_2}  +
\frac{g_2}{n+g_1+g_2}\frac{n+g_2}{n+g_1+g_2}
\right) \frac{1}{k+2} +\right.\\\left.
\frac{g_1}{n+g_1+g_2}\frac{g_2}{n+g_1+g_2}\frac{1}{k+3}
\right].
\end{multline*}
Using the reasoning above, we see that,
\begin{equation*}
\frac{n+g_1}{n+g_1+g_2}\frac{n+g_2}{n+g_1+g_2}\frac{1}{k+1},
\end{equation*}
gives the probability of eating when a random member of the free group collides with no members from group 1 or group 2 and $k$ additional members from his own group. The term,
\begin{equation*}
\left(
\frac{g_1}{n+g_1+g_2}\frac{n+g_1}{n+g_1+g_2}  +
\frac{g_2}{n+g_1+g_2}\frac{n+g_2}{n+g_1+g_2}
\right) \frac{1}{k+2},
\end{equation*}
gives the probability that a member of the free group eats if he collides with exactly one member from either group 1 or group 2 and $k$ other members from his group. Finally, 
\begin{equation*}
\frac{g_1}{n+g_1+g_2}\frac{g_2}{n+g_1+g_2}\frac{1}{k+3},
\end{equation*}
gives the probability that a member of the free group eats if he collides with $k$ other members of his group and a member from group 1 and a member of group 2. 

It is clear that the structure of \cref{eqn:p,eqn:q} can be generalised to an arbitrary number of groups, as noted before, the implications of which, we consider for future work. Suppose we have $g_i = \alpha_i n$ for $i \in \{1,2\}$ and generalising from \cite{HBG2023}, we define 
\begin{equation*}
    \beta_i = \frac{g_i}{g_1+g_2+n} = \frac{\alpha_i n}{n+\alpha_1n + \alpha_2n} = \frac{\alpha_i}{1+\alpha_1 + \alpha_2},
\end{equation*}
to be the proportion of the population in dining club $i$. Then as $n \to \infty$, we can compute,
\begin{equation*}
    p(\beta_1,\beta_2) = \frac{(1-\beta_1-2\beta_2) + e^{\beta_1 + \beta_2 - 1}\left[\beta_1(1-\beta_2) + \beta_2(3-\beta_2) - 1\right]}{(1 - \beta_1 - \beta_2)^2},
\end{equation*}
as the asymptotic probability that a member of group 1 eats with $p(\beta_2,\beta_1)$ being the asymptotic probability that a member of group 2 eats. Let $q(\beta_1,\beta_2)$ denote the probability that an arbitrary member of the free group eats. We will compute this using the following the reasoning from \cite{HBG2023}, let $S_i$ be a Bernoulli random variable taking the value $1$ if a member of group i ($i\in\{1,2\}$) eats and $0$ otherwise. For notational consistency, let $S_0$ be a Bernoulli random variable taking the value $1$ if a member of the free group eats and finally, let $S$ be a Bernoulli random variable taking the value $1$ just in case an arbitrary member of the population eats. Then,
\begin{align*}
&\mean{S_1} = p(\beta_1,\beta_2)\\
&\mean{S_2} = p(\beta_2,\beta_1)\\
&\mean{S} = \beta_1 p(\beta_1,\beta_2) + \beta_2p(\beta_2,\beta_1) + (1-\beta_1-\beta_2)q(\beta_1,\beta_2) = \\
&\hspace*{20em}1-e^{\beta _1+\beta _2-1} \left(1 - \beta_1\right) \left(1 - \beta_2\right).
\end{align*}
From this, we can express $q(\beta_1,\beta_2)$ compactly as,
\begin{equation*}
\mean{S_0} = q(\beta_1,\beta_2) = \frac{1-e^{\beta _1+\beta _2-1} \left(1 - \beta_1\right) \left(1 - \beta_2\right) - \beta_1p(\beta_1,\beta_2) - \beta_2p(\beta_2,\beta_1)}{1 - \beta_1 - \beta_2}.
\end{equation*}

We note that $\mean{S_i}$ can also be thought of as the mean meal size of an individual in group $i$. As such, we can think of $\mean{S_i}$ as the fitness for an individual in group $i$. Now assume the population evolves according to the following microscopic rules. If two individuals meet, they will compare their expected meal size. If the difference is greater than some tolerance $\tau > 0$, then the individual with smaller expected meal size will switch to the group of the individual with larger expected meal size. This describes a group ``stickiness'' with an individual only willing to leave a group if it is likely they will increase their fitness sufficiently. From a biological perspective, we may treat the groups as species. In that case, this dynamic occurs only if the fitness difference between two competing species is sufficiently large to ensure successful reproduction. To model this scenario, we use a variation on a compartmented model \cite{H00}. Consider a fixed (or infinite population) and let $\mathbf{p} = \langle{p_1,\dots, p_n}\rangle$ be a vector of species (group) proportions, then,
\begin{equation*}
\frac{d p_i}{dt}  
= p_i \left\{\sum_{j}\left[G_{ji}(\mathbf{p}) - G_{ij}\,(\mathbf{p})\right]p_j\right\},
\label{eqn:DSYS}
\end{equation*}
where,
\begin{equation}
G_{ij}(\mathbf{p})= \sigma\left[F_j(\mathbf{p}) - F_i(\mathbf{p})\right].
\end{equation}
Here $F_i(\mathbf{p})$ is the (mean) fitness of species (group) $i$ and $G_{ij}(\mathbf{p})$ can be thought of as a transition probability (in the sense of a Markov chain). The function $\sigma:\mathbb{R} \rightarrow [0,1]$ is a non-negative monotone increasing transition function, that defines, quantitatively, the character of compartment switching. Just as in \cite{EGB16},  the value $\sigma(0)$ represents an equilibrium when the inflow and outflow are equal with smooth (i.e., less sharp) transitions around $0$ representing a ``higher temperature'' (in a spin glass sense). For the sequel, we assume that $\sigma$ is a sigmoid in which the population drifts toward better-performing strategies. 
 
\begin{figure}
\centering
\includegraphics[width=0.6\textwidth]{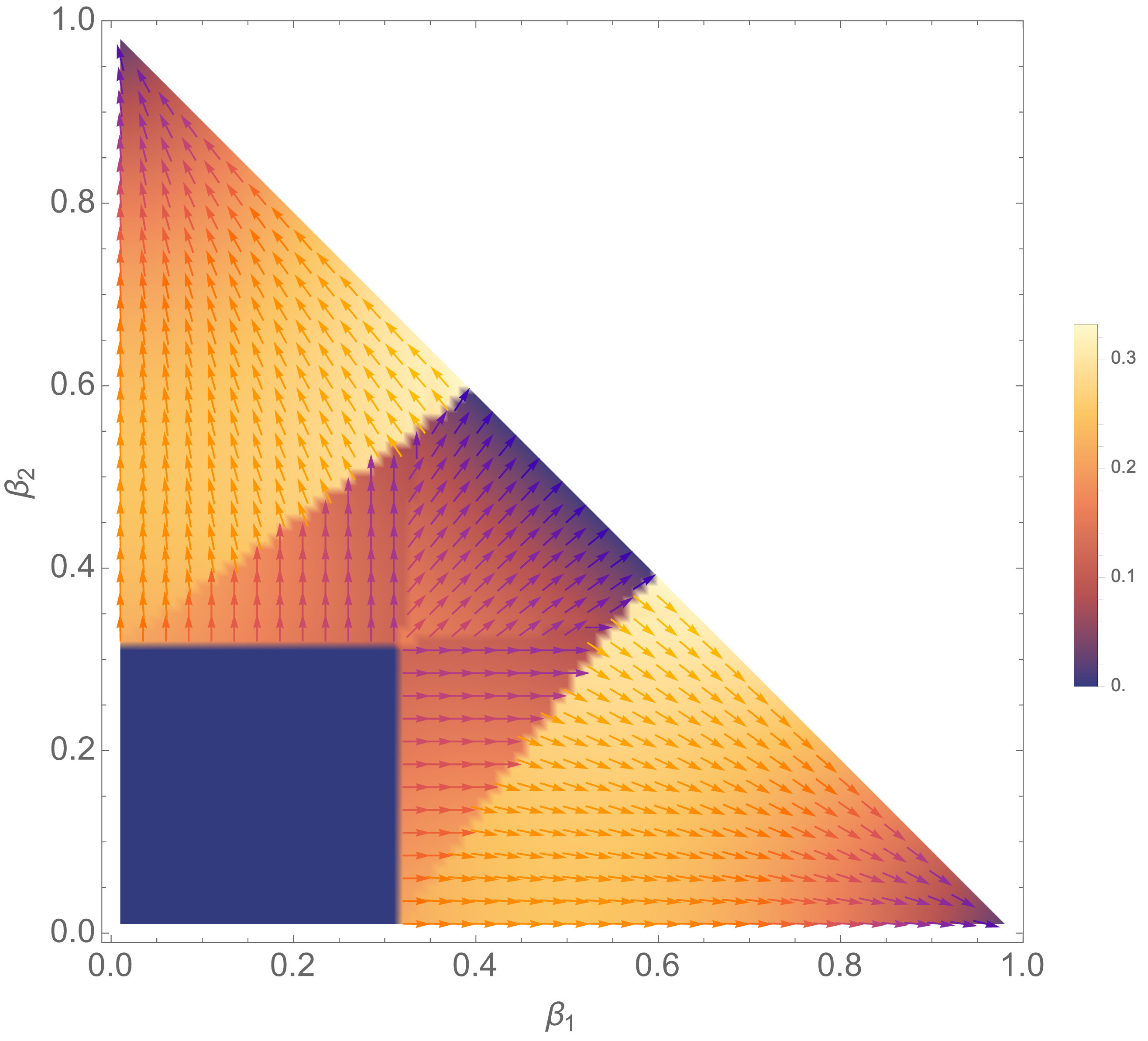}
\caption{Vector field plot for the dynamics when $\tau = 0.1$. The colouring shows the norm of the vectors in the field.}
\label{fig:TwoGroupField}
\end{figure}

Applying this dynamic to our model yields the coupled system of differential equations,
\begin{align*}
\dot{\beta}_1 &= \beta_1(1-\beta_1-\beta_2)\sigma\left(\mean{S_1} - \mean{S_0}\right)+ \beta_1\beta_2\sigma\left(\mean{S_1} - \mean{S_2}\right) -\\
&\hspace*{10em}\beta_1  \left[
(1-\beta_1-\beta_2)\sigma\left(\mean{S_0} - \mean{S_1}\right) + 
\beta_2\sigma\left(\mean{S_2} - \mean{S_1}\right)
\right]\\
\dot{\beta}_2 &= \beta_2(1-\beta_1-\beta_2)\sigma\left(\mean{S_2} - \mean{S_0}\right)+ \beta_2\beta_1\sigma\left(\mean{S_2} - \mean{S_1}\right) -\\
&\hspace*{10em}\beta_2  \left[
(1-\beta_1-\beta_2)\sigma\left(\mean{S_0} - \mean{S_2}\right) + 
\beta_1\sigma\left(\mean{S_1} - \mean{S_2}\right)
\right].
\end{align*}
The described dynamics are recovered precisely if,
\begin{equation*}
\sigma(z) = \begin{cases}
    1 & \text{if $z \geq \tau$}\\
    0 & \text{otherwise}.
\end{cases}
\end{equation*}
For the remainder of this section, we assume $\sigma(z)$ has this form. An example vector field plot for the resulting dynamic system is shown in \cref{fig:TwoGroupField} for $\tau = 0.1$.
\cref{fig:TwoGroupField} illustrates a general property of the model that monocultures are fixed points. This is easy to see for $\beta_1 = \beta_2 = 0$. The more interesting case occurs at the other monocultures $\beta_1 = 1$ or, $\beta_2 = 1$, and it is instructive to analyse the evolution under the constraint $\beta_1 + \beta_2 = 1$ (i.e., there are no free agents in the population). To obtain the dynamics in this case, substitute $\beta_2 = (1 - \epsilon) - \beta_1$ into the expressions for $\mean{S_1}$ and $\mean{S}_2$ and let $\epsilon \to 0$. The dynamics of $\beta_1$ become,
\begin{equation*}
    \dot{\beta}_1 = \left(1-\beta_1\right) \beta_1 \left[\sigma\left(\beta_1-\frac{1}{2}\right) - \sigma\left(\frac{1}{2}-\beta_1\right)\right].
\end{equation*}
A bit of additional computation shows that $\dot{\beta}_2 = -\dot{\beta}_1$, as expected. Letting $\beta_1 = 1$, $\dot{\beta}_1 = 0$, thus proving that all monocultures are fixed points by symmetry. Moreover, for the infinite population case, we have also shown that when there are no agents in the free group, we must have,
\begin{equation*}
    \beta_1(0) \geq \frac{1}{2} + \tau \quad \text{or} \quad \beta_2(0) \geq \frac{1}{2} + \tau,
\end{equation*}
for the growth or decay in $\beta_1$. When $\beta_1(0) \geq \frac{1}{2} + \tau$, then $\beta_1$ exhibits logistic growth. When $\beta_1(0) <\frac{1}{2} + \tau$, $\beta_1$ exhibits logistic decay. 

We note that equal redistribution of food in the infinite population case will not change $\mean{S_i}$, as it simply ensures that all individuals receive the same quantity of food, as pointed out in \cite{HBG2023}. However, this redistribution cannot change the mean as the probability that an arbitrary agent is selected for food, (and then shares it) remains the same. Thus, redistribution in the infinite case does not change the resulting dynamics in the presence of tolerance in just the same way that it does not affect the dynamics in a single dining club scenario as pointed out in \cite{HBG2023}.

In the remainder of this paper, we focus on the case where the population has been divided into two dining clubs. That is, there are no free agents. As we show in our experimental results, the infinite population analysis only captures the behaviour assuming an exceptionally large population using perfect information about $\mean{S_1}$ and $\mean{S_2}$. In finite population simulations (as in the real world), individuals make decisions based on a finite quantity of information, leading to more interesting dynamics.

\section{Simulation Functions}\label{sec:SimFunctions}
We simulate a finite population of agents to determine how closely the resulting dynamics adhere to the theoretical dynamics derived above. In these simulations, we use three sub-routines. Let $\mathcal{G}_1$, $\mathcal{G}_2$, and $\mathcal{G}_0$ be sets denoting the populations in dining club one and dining club two and the population in neither club. 

\paragraph*{Monte Carlo Restaurant Assignment} The \texttt{MonteCarlo} subroutine models the action of agents randomly choosing restaurants and is shown below.
\begin{algorithmic}
\STATE \texttt{MonteCarlo:}
\STATE \textbf{Input:} $\mathcal{G}_0$, $\mathcal{G}_1$, $\mathcal{G}_2$
\STATE $N \gets |\mathcal{G}_0| + |\mathcal{G}_1| + |\mathcal{G}_2|$
\FOR{$i \in 1,2$}
\STATE $R = \{1,\dots,N\}$ \COMMENT{This is the restaurant set.}
\FOR{$j \in \mathcal{G}_i$}
\COMMENT{Assign restaurants to club members.}
\STATE Select $k$ randomly from $R$.
\STATE Assign  $j$ to $k$ and remove $k$ from $R$.
\ENDFOR
\ENDFOR
\FOR{$j \in \mathcal{G}_0$}
\COMMENT{Assign restaurants to non-club agents.}
\STATE Select $k$ randomly from $R$.
\STATE Assign $j$ to $k$.
\ENDFOR
\FOR{$k \in R$}
\STATE Randomly feed (give utility $1$) to an agent assigned to restaurant $k$.
\STATE All other agents assigned to restaurant $k$ do not eat (receive utility $0$).
\ENDFOR
\STATE \textbf{Output:} Utility functions $u_i:\mathcal{G}_i \to \{0,1\}$ and $u_0:\mathcal{F}\to\{0,1\}$.
\end{algorithmic}
This subroutine is used to calculate the utility of an agent in a group. Within a group, each agent is assigned randomly to a unique restaurant to ensure that no two agents in the same group are assigned to the same restaurants. Then for all the restaurants that have agents assigned, we randomly give agents food, i.e., utility $1$. All other agents are assigned no food (utility $0$).

The \texttt{MonteCarlo} subroutine can be run $M > 0$ times and the average utility for each agent can be calculated and used in subsequent subroutines. That is, let $u_i^l:\mathcal{G}_i \to \{0,1\}$ be the utility function for Group $i$ on the $l^\text{th}$ run ($l\in\{1,\dots,M\})$. Then, for $k \in \mathcal{G}_i$,
\begin{equation}
    u_i(k) = \frac{1}{M}\sum_{l} u_i^l(k),
    \label{eqn:EstUtility}
\end{equation}
is the proportion of times agent $k$ in Group $i$ eats over $M$ runs. A similar proportion can be computed for agents in $\mathcal{G}_0$.

\paragraph*{Agent Reshuffling}
We use a \texttt{Reshuffle} subroutine to move agents among groups based on their probability of eating and their tolerance to differences in those probabilities. The algorithm is shown below. 
\begin{algorithmic}
\STATE \texttt{Reshuffle:}
\STATE \textbf{Input:} $\mathcal{F}$, $\mathcal{G}_1$, $\mathcal{G}_2$, $\tau$
\STATE Set $\mathcal{P} = \mathcal{F} \cup \mathcal{G}_1 \cup \mathcal{G}_2$
\WHILE{There is more than 1 agent in the list $\mathcal{P}$}
\STATE Choose two agents $i$ and $j$ at random from $\mathcal{P}$.
\STATE Let $\texttt{Group}(i)$ (resp. $\texttt{Group}(j)$) be the group to which $i$ (resp. $j$) belongs. Note: $i$ and $j$ can belong to any group and are chosen at random.
\STATE Let $p_i$ (resp. $p_j$) be the probability that $i$ (resp. $j$) eats. \COMMENT{There are utilities as computed above in \cref{eqn:EstUtility}.}
\IF{$p_i > p_j + \tau$}
\STATE Move $j$ to $\texttt{Group}(i)$
\ELSIF{$p_j > p_i + \tau $}
\STATE Move $i$ to $\texttt{Group}(j)$
\ENDIF
\ENDWHILE
\end{algorithmic}
Two agents are selected randomly from the pool of all agents. If their relative utility differs by an amount greater than $\tau$, the agent with lower utility joins the group of the agent with higher utility. 

\paragraph{Utility Redistribution and Freeloading}
When we study the effect of redistribution of resources in groups, we use the \texttt{Redistribute} subroutine that sets the utility of all agents in a group to the mean utility of the group. Given a utility function $u_i$ computed from the \texttt{MonteCarlo}, the \texttt{Redistribute} subroutine is shown below.
\begin{algorithmic}
\STATE \texttt{Redistribute:}
\STATE \textbf{Input:} $\mathcal{G}_i$, $u_i^l$ 
\STATE Let $g_i = |\mathcal{G}_i|$
    \FOR{$k \in \mathcal{G}_i$}
        \STATE  $\bar{u}_i^l(k) = \frac{1}{g_i}\sum_j u_i^l(k)$
    \ENDFOR
\end{algorithmic}
This is consistent with the redistribution mechanism studied in \cite{HBG2023}. The utility received by each agent is then the group mean utility derived from competing for service at a restaurant.

We can modify this algorithm to incorporate randomized freeloading from an external group $\mathcal{G}_{i'}$.
\begin{algorithmic}
\STATE \texttt{Redistribute-Freeload:}
\STATE \textbf{Input:} $\mathcal{G}_i$, $u_i^l$, $\mathcal{G}_{i'}$ , $u_{i'}^l$, $p_\text{freeload}$
\STATE Let $g_i = \vert \mathcal{G}_i \vert$
\STATE Let $g_{i'} = \vert \mathcal{G}_{i'} \vert$
\STATE Let $Z_{i'} = \{k \in \mathbb{Z} : u_{i'}^l(k) = 0\}$.
\STATE Let $z_{i'} = \vert Z_{i'} \vert$.
    \FOR{$k \in \mathcal{G}_i$}
        \STATE  $\bar{u}_i^l(k) = \frac{1}{g_i + z_{i'}}\sum_j u_i$
    \ENDFOR
    \FOR{$k \in \mathcal{G}_{i'}$}
        \IF{$k \in Z_{i'}$ \textbf{and} $\text{\texttt{random}}()<p_\text{freeload}$}
            \STATE $\bar{u}_{i'}^l(k) = \frac{1}{g_i + z_{i'}}\sum_j u_i$
        \ELSE 
            \STATE $\bar{u}_{i'}^l(k) = u_{i'}^l(k)$
        \ENDIF
    \ENDFOR
\end{algorithmic}
In this case, the redistributing group is parasitized by a second group at random. When $p_\text{freeload} = 1$, then all individuals who do not eat in the second ground share the resources of the redistributing group. Here, $\text{\texttt{random}}()$ indicates a uniform random number in $[0,1]$.

\section{Experimental Results without Redistribution}\label{sec:NoRedist}
In our first experiment, we ran a simple KPRP game between two dining clubs without any free agents or redistribution among members of the dining clubs. The experimental procedure is shown below.
\begin{algorithmic}
\STATE \texttt{Reshuffling Experiment}
\FOR{$t\in\{1,\dots,T\}$}
\FOR{$l\in\{1,\dots,M\}$}
\STATE \texttt{MonteCarlo} 
\ENDFOR
\STATE \texttt{Reshuffle}
\ENDFOR
\end{algorithmic}

In our experiments, we used $N\in\{100,200\}$ agents, initially divided equally between the two clubs. Agents could switch between clubs using reshuffling. We used Monte Carlo assignment 100 times (i.e., $M = 100$) to approximate individual agent mean utility (probability of eating) before reshuffling. In each experiment, tolerance was allowed to range from $0$ to $0.3$ increasing by $0.01$. We ended each run when a dominant group formed (i.e., $M$ agents were present in one group) or after $T=100$ reshuffles. We performed 30 replications of this process with populations of both $N = 100$ and $N = 200$ agents. 

\begin{figure}[htbp]
\centering
\includegraphics[width=0.45\textwidth]{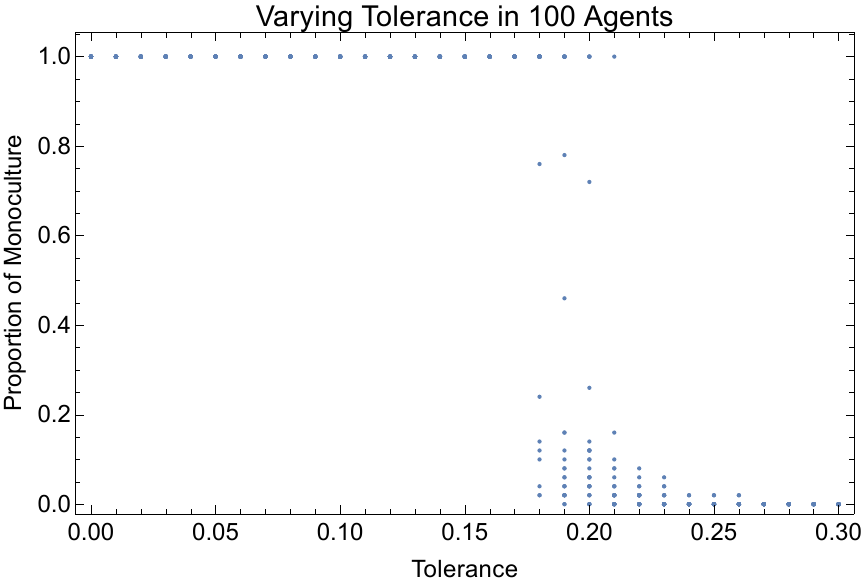} \quad 
\includegraphics[width=0.45\textwidth]{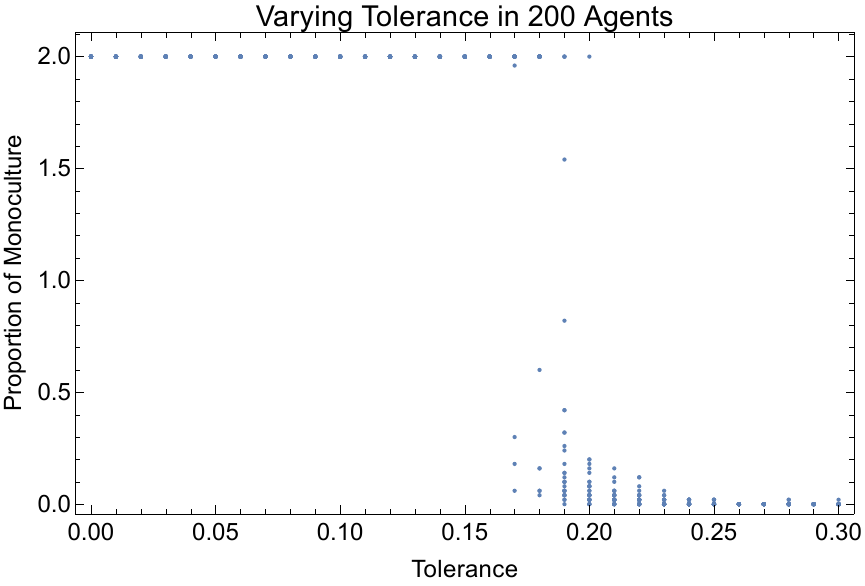}
\caption{(Left) Per capita absolute difference in groups sizes for a population of 100 agents as a function of tolerance. (Right)  Per capita absolute difference in groups sizes for a population of 200 agents as a function of tolerance. Note both experiments show a phase change at $\tau \approx 0.185$.}
\label{fig:ExperimentalResult1}
\end{figure}

Experimental results are shown in \cref{fig:ExperimentalResult1}. We plot the per capita absolute difference in group size,
\begin{equation*}
    \frac{\left\vert \vert\mathcal{G}_1\vert - \vert\mathcal{G}_2\vert \right\vert}{N},
\end{equation*}
which estimates the probability that the system evolves to a group monoculture on or before 100 reshuffles.
The results suggest a sharp phase change at $\tau \approx 0.185$. At this point, a dominant group does not form after 100 reshuffles, and the population stays mixed between the two groups. 

\begin{figure}[htbp]
\centering
\includegraphics[width=0.65\textwidth]{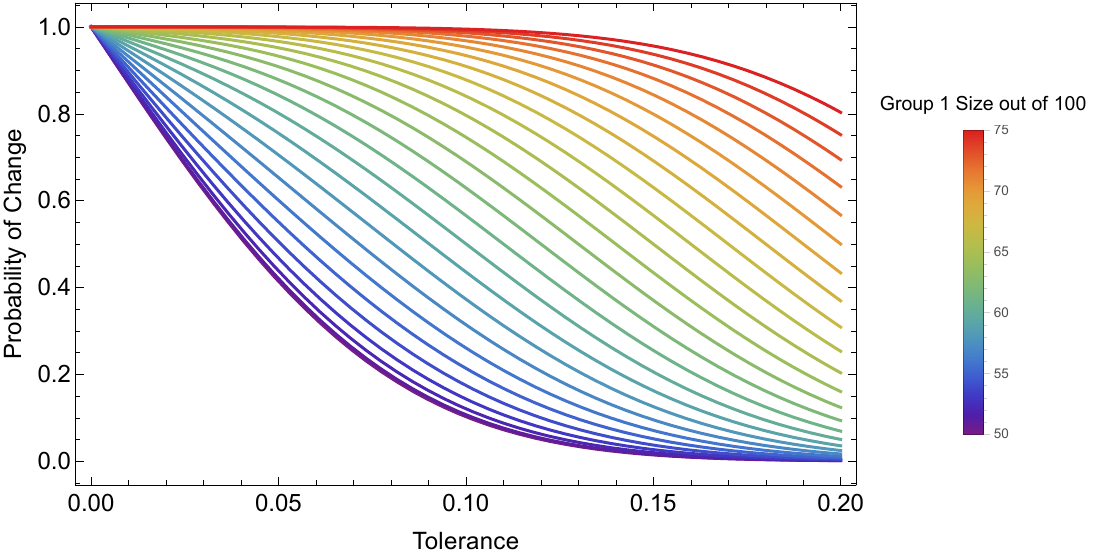}
\caption{The probability of moving from a state $(g_1,g_2)$ to $(g_1\pm 1,g_2\mp1)$ for varying values of $g_1$ and $g_2$ assuming $g_1 + g_2 = 100$.}
\label{fig:ProbChange}
\end{figure}

\begin{figure}[htbp]
\centering
\includegraphics[width=0.45\textwidth]{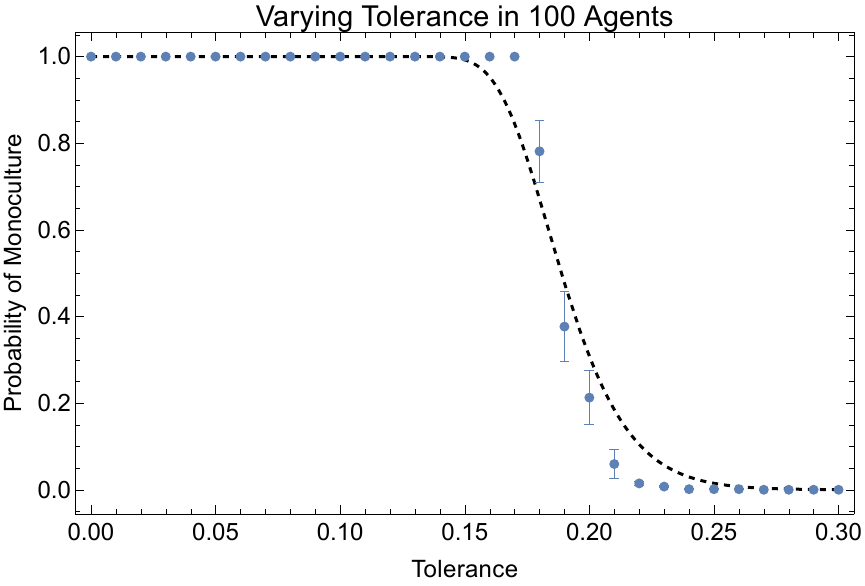}
\quad
\includegraphics[width=0.45\textwidth]{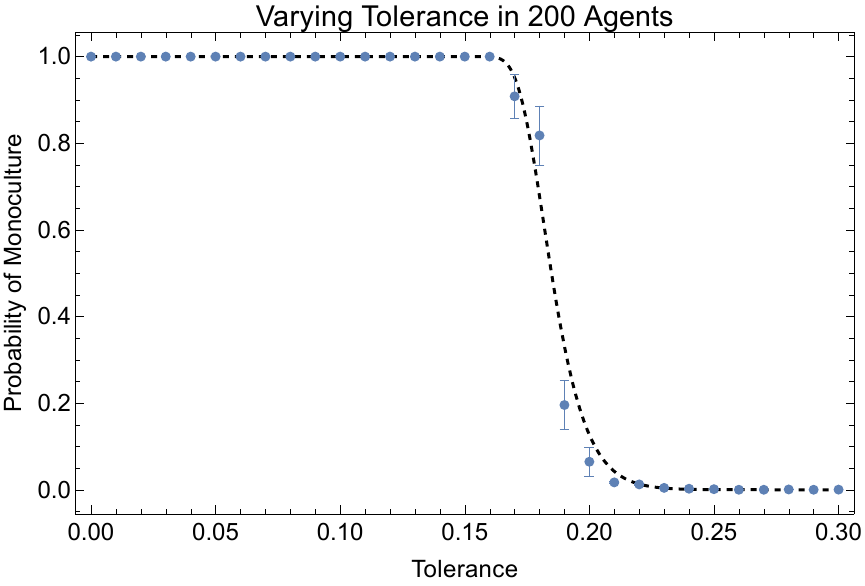}
\caption{Probability of the formation of a monoculture in populations of size 100 (left) and (200) right with fitted estimator.}
\label{fig:ProbChangeModel}
\end{figure}

Using \cref{eqn:p} and setting $n = 0$, we can explain these results for finite populations. We have,
\begin{equation*}
    p(0,g_1,g_2) = \tilde{p}(g_1,g_2) = \frac{1}{2} \left(\frac{g_1}{g_1+g_2}+1\right).
\end{equation*}
Then when two members of opposite groups meet at random, they are comparing two binomial random variables,
\begin{equation*}
    S_1 \sim \mathrm{Binom}\left[m,\tilde{p}(g_1,g_2)\right],
\end{equation*}
and
\begin{equation*}
    S_2 \sim \mathrm{Binom}\left[m,\tilde{p}(g_2,g_1)\right].
\end{equation*}
Assume $m \gg 1$ (in our experiments, we use $m = 100$). Let,
\begin{equation*}
    \tilde{\mu} = m\left[\tilde{p}(g_1,g_2) - \tilde{p}(g_2,g_1)\right]
\end{equation*}
and
\begin{equation*}
    \tilde{\sigma}^2 = m\tilde{p}(g_1,g_2)\left[1 - \tilde{p}(g_1,g_2)\right] + m\tilde{p}(g_2,g_1)\left[1 - \tilde{p}(g_2,g_1)\right].
\end{equation*}
Then,
\begin{equation*}
    S_1 - S_2 \sim \mathcal{N}\left(\tilde{\mu},\tilde{\sigma}^2\right).
\end{equation*}
From this, we have,
\begin{multline*}
    p_\Delta(g_1,g_2) = \mathrm{Pr}\left(\left\lvert{S_1 - S_2}\right\rvert > \tau\right) \approx \\
    1 - \frac{1}{2} \left[\mathrm{erf}\left(\frac{m \left(-2 g_1+2 n \tau +n\right)}{\sqrt{2} n
   \sqrt{\frac{2 g_1 g_2 m}{n^2}+m}}\right)+\mathrm{erf}\left(\frac{m \left(2 g_1+n (2
   \tau -1)\right)}{\sqrt{2} n \sqrt{\frac{2 g_1 g_2 m}{n^2}+m}}\right)\right].
\end{multline*}
This is the probability of randomly increasing one group's size as a result of one random comparison. When $g_1 = g_2$, this simplifies to,
\begin{equation*}
    p_\Delta\left(\tfrac{n}{2},\tfrac{n}{2}\right) = \mathrm{Pr}\left(\left\lvert{S_1 - S_2}\right\rvert > \tau \vert g_1 = g_2 = \tfrac{n}{2}\right) = 1 - \mathrm{erf}\left(\frac{2 \sqrt{m} \tau }{\sqrt{3}}\right)
\end{equation*}
As $g_1$ (resp. $g_2$) increases, the probability that this group will become dominant (i.e., all players will join that group) increases. This is easiest to see by plotting $p_\Delta(g_1,g_2)$ for varying values of $g_1,g_2$ so that $g_1 + g_2 = N$ is fixed. This is shown in \cref{fig:ProbChange}.
Thus, we conclude that for the finite population, the dynamics of $(g_1,g_2)$ are governed by a discrete Markov chain with state space,
\begin{equation*}
    Q = \left\{(g_1,g_2) : g_1 + g_2 = n, g_1,g_2 \in \mathbb{Z}_+\right\},
\end{equation*}
and with transitions defined from each state to every other state. From \cref{fig:ProbChange}, we know that (e.g.) once an initial move is made away from an initially balanced condition $g_1 = g_2 = \tfrac{N}{2}$, the more likely the system will settle in a state with only one group. Thus, we can approximate the probability that the system will reach a grand coalition (a single dining club with all players) as,
\begin{equation*}
    p^*(\tau) \approx 1 - \mathrm{erf}\left(\frac{2 \sqrt{m} \tau }{\sqrt{3}}\right)^{r},
\end{equation*}
where $r \approx 337$ is determined by statistical fit for $N=100$ and $r \approx 35,113$ for $N = 200$. This is (very roughly) the probability that the system transitions away from the initial condition with $g_1 = g_2 = \tfrac{n}{2}$ as a result of any pair of player comparisons over several epochs. The value of $r$ is a result of the approximation. The model and experimental data are shown in \cref{fig:ProbChangeModel}.
Error bars are computed using the asymptotic formula for the confidence around the mean of a Bernoulli distribution, 
\begin{equation}
    \pm Z_{1-\alpha/2}\sqrt{\frac{p(1-p)}{n}}.
    \label{eqn:CI}
\end{equation}
Here $n = 30$, the number of replications.

To compute exact behaviour near the phase transition would require analysis of the Markov chain governing the finite population system. As we have already shown, as $N \to \infty$, the system will not transition from the limiting initial condition $\beta_1 = \beta_2 = \tfrac{1}{2}$, which is a stable equilibrium if $\tau > 0$ and unstable when $\tau = 0$.

\section{Experimental Results with Redistribution and Freeloading}\label{sec:Redistribution}
We modified the experimental procedure from the preceding section to include redistribution in one or more groups and/or freeloading by one group on another. 
\begin{algorithmic}
\STATE \texttt{Redistribution and Freeloading Experiment}
\FOR{$t\in\{1,\dots,T\}$}
\FOR{$l\in\{1,\dots,M\}$}
\STATE \texttt{MonteCarlo} 
\IF{Only $\mathcal{G}_1$ Redistributes}
\STATE \texttt{Redistribute}$(\mathcal{G}_1,u_1)$
\IF{$\mathcal{G}_2$ Freeloads}
\STATE \texttt{Redistribute-Freeload}$(\mathcal{G}_1,u_1,\mathcal{G}_2,u_2)$
\ENDIF
\ELSE
\STATE \texttt{Redistribute}$(\mathcal{G}_1,u_1)$
\STATE \texttt{Redistribute}$(\mathcal{G}_2,u_2)$\COMMENT{Both groups redistribute}
\ENDIF
\ENDFOR
\STATE \texttt{Reshuffle}
\ENDFOR
\end{algorithmic}
As before, we used $N\in\{100,200\}$ agents, initially divided equally between the two clubs. Agents could switch between clubs using reshuffling. We used Monte Carlo assignment 100 times (i.e., $M = 100$) to approximate individual agent mean utility (probability of eating) and then either performed redistribution or redistribution with freeloading. In each experiment, tolerance was allowed to range from $0$ to $0.3$ increasing by $0.01$. We ended each run when a dominant group formed (i.e., $M$ agents were present in one group) or after $T=100$ reshuffles. We performed 30 replications of this process with populations of both $N = 100$ and $N = 200$ agents for experiments involving redistribution only. For freeloading, we performed 20 replications.

\begin{figure}[htbp]
\centering
\includegraphics[width=0.45\textwidth]{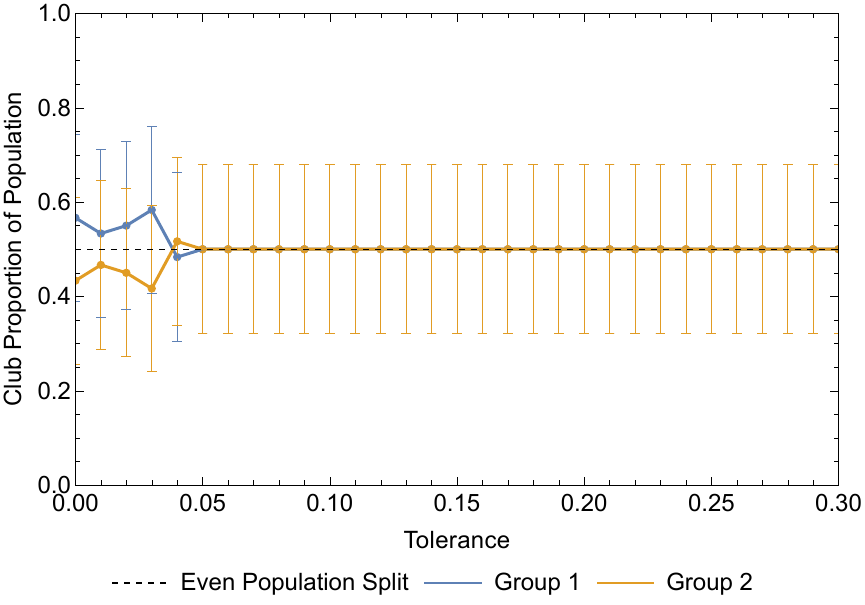} \quad
\includegraphics[width=0.45\textwidth]{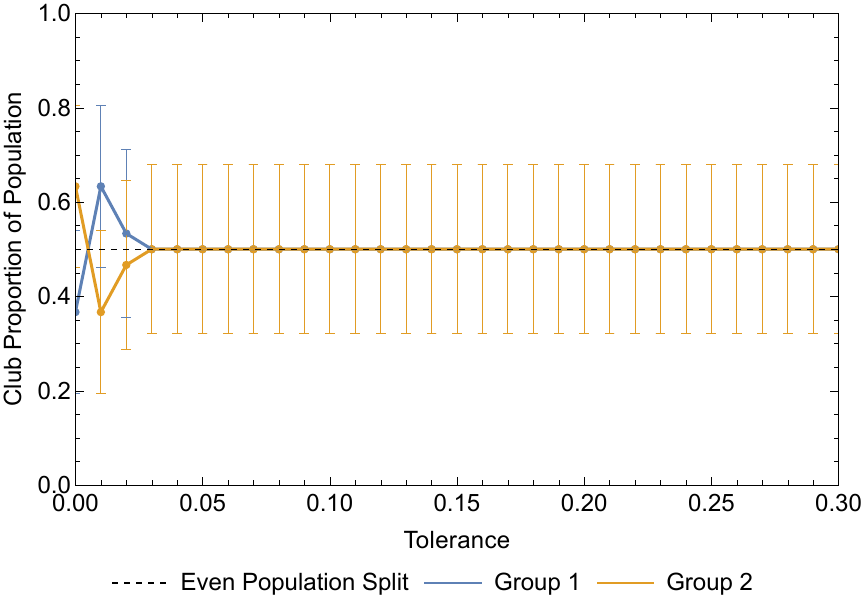}
\caption{(Left) Population size 100 with both groups redistributing resources. (Right) Population size 200 with both groups redistributing resources.}
\label{fig:TwoGroupsRedistribute}
\end{figure}

When both groups redistribute food, the resulting dynamics are closer to the expected behaviour for a large (infinite) population. That is, a dining club monoculture fails to form for much lower tolerances, as shown in \cref{fig:TwoGroupsRedistribute}. 
As before, error bars are computed using \cref{eqn:CI}. The observed behaviour is the result of the law of large numbers, where each distribution of the agents' utilities is replaced by a single value (for each group). This value is itself approximately normally distributed as,
\begin{equation*}
    u_i \sim \mathcal{N}(p_i, p_i(1-p_i)/g_i).
\end{equation*}
We note, however, that at any given comparison time (i.e., when \texttt{SwapGroups} is executed), every agent in each group shares a common value. Thus, on average, there is no flow from one group to another. We will see in the sequel that this approximation does not always explain observed behaviour in the case of asymmetric redistribution. 

\begin{figure}[htbp]
\centering
\includegraphics[width=0.45\textwidth]{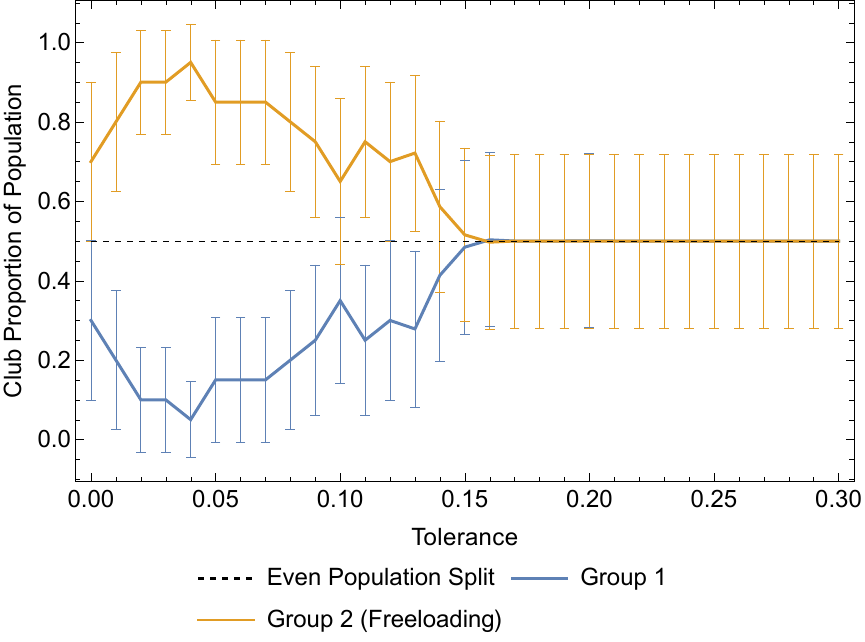} \quad 
\includegraphics[width=0.45\textwidth]{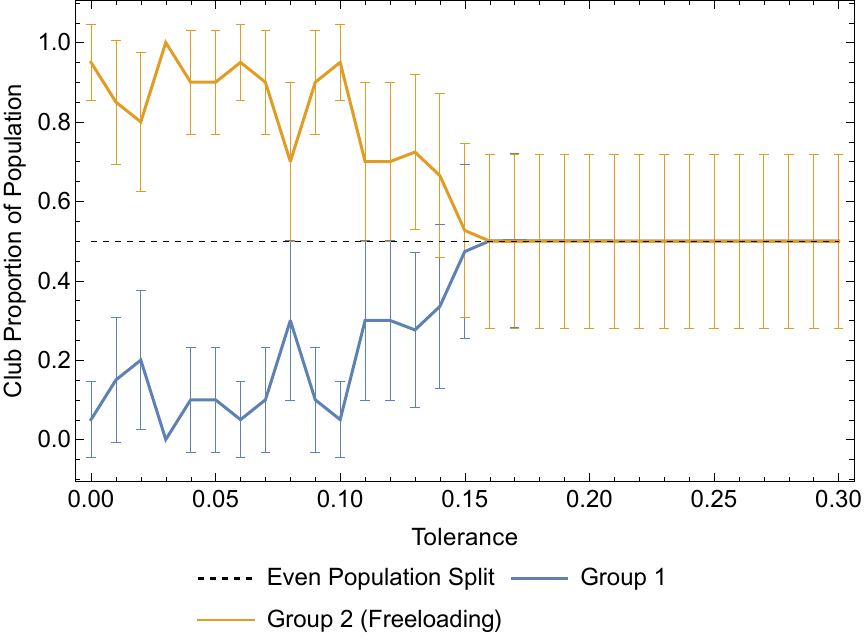}\\
\includegraphics[width=0.45\textwidth]{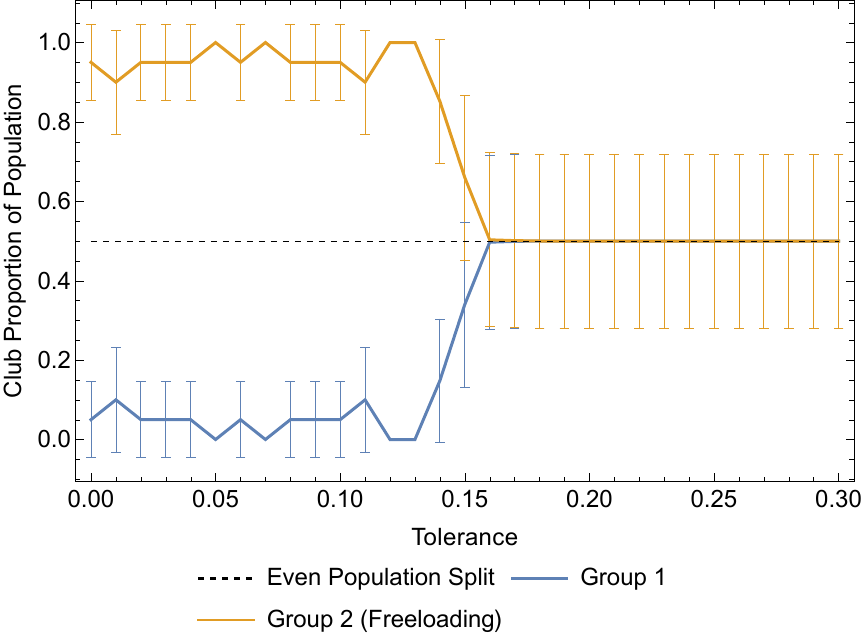} \quad 
\includegraphics[width=0.45\textwidth]{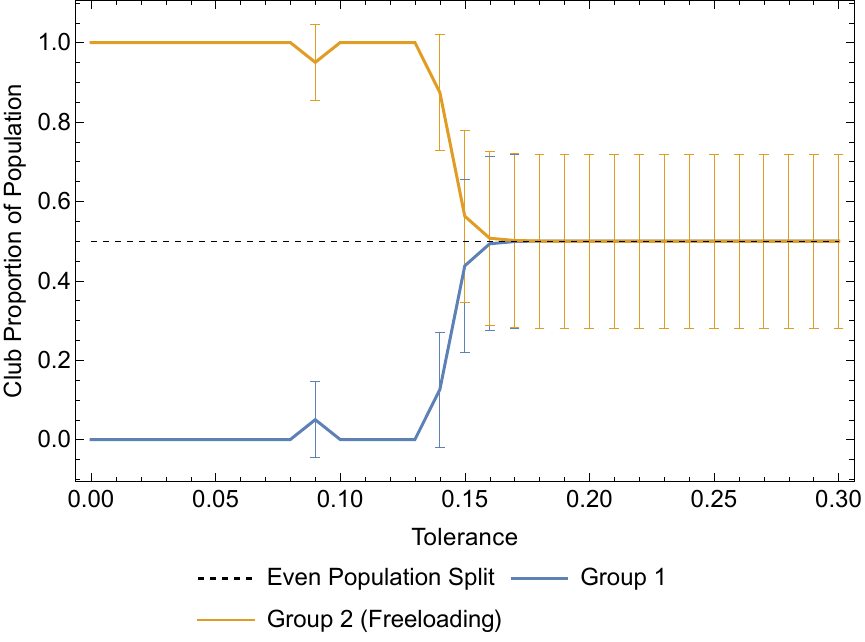}
\caption{(Top) Group 1 redistributes food to its members and freeloading members of Group 2. Members of group 2 freeload with probability $0.025$. (Bottom)  Group 1 redistributes food to its members and freeloading members of Group 2. Members of group 2 freeload with probability $0.025$. (Left) Population size is 100. (Right) Population size is 200.}
\label{fig:Freeloading}
\end{figure}

When one group redistributes and the other group is allowed to freeload off the distributions, the redistributing group frequently collapses at small tolerances, even if the probability that an agent from the other dining club freeloads is small. This is illustrated in \cref{fig:Freeloading}.
Plots in \cref{fig:Freeloading} (top) show the impact of freeloading with probability 2.5\%. Even at this low rate, Group 1 is more frequently driven to extinction (out-competed) by Group 2. Plots in \cref{fig:Freeloading} (bottom) show the impact of freeloading with a probability of 5\%, which has a more pronounced effect on Group 1. The plots on the left both have 100 agents in the population. Plots on the right have 200 agents in the population. In this case, $n = 20$ (replications) for computing the error bars using \cref{eqn:CI}. These results are not surprising, given prior work in \cite{HBG2023} that showed freeloading by non-dining club agents can destabilize a group. 

\begin{figure}[htbp]
\centering
\includegraphics[width=0.45\textwidth]{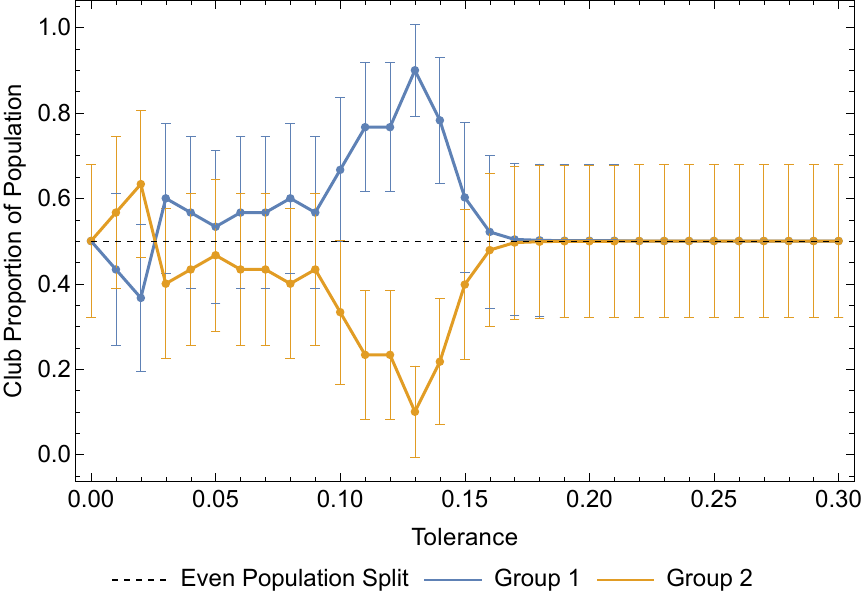}\quad
\includegraphics[width=0.45\textwidth]{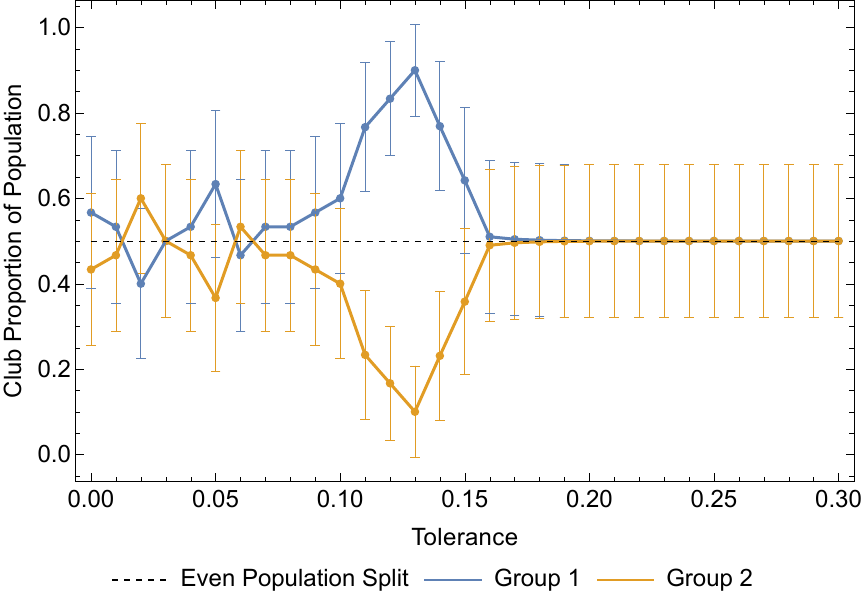}
\caption{(Left) Proportion of population of each dining club when group one redistributes and group two does not. (Left) Population size is 100. (Right) Population size is 200.}
\label{fig:OneRedistribute}
\end{figure}

\begin{figure}[htbp]
\centering
\includegraphics[width=0.45\textwidth]{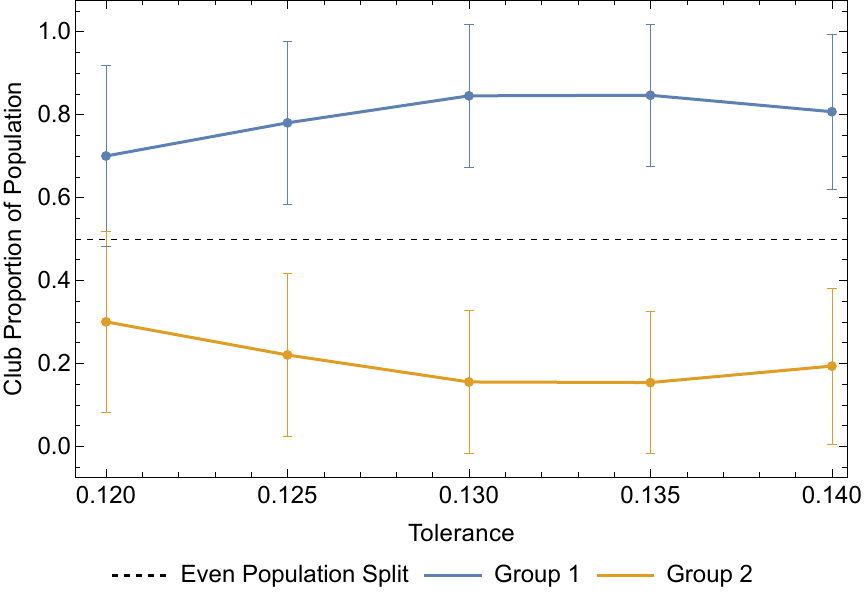} \quad
\includegraphics[width=0.45\textwidth]{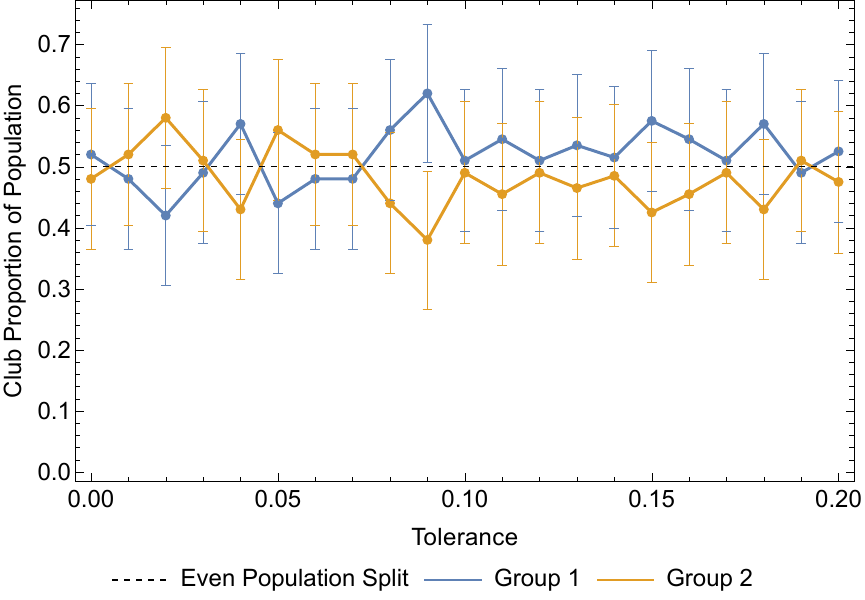}
\caption{(Left) Higher resolution simulation results with Group 1 redistributing resources and Group 2 not redistributing resources. We used 100 replications and error bars correspond to a $p$-value of $10^{-6}$, showing there is a high probability of a difference between the two groups. (Right) Simulation with Group 1 redistributing resources and Group 2 not redistributing resources with a population of only four agents. Error bars correspond to a $p$-value of $0.01$.}
\label{fig:DetailsRedistribution}
\end{figure}

More interesting behaviour emerges when only one group redistributes resources (food) and freeloading is prohibited. We noted in \cref{sec:MathModel} that for infinite populations, redistribution does not affect the phase space of the dynamical system (e.g., \cref{fig:TwoGroupField}). Experimental results shown in \cref{fig:OneRedistribute} show this is not the case.
The data show that in the finite population case, the group engaging in resource redistribution has a slight advantage when $0.12 \lesssim \tau \lesssim 0.14$. This holds for populations of size 100 and 200 (\cref{fig:OneRedistribute} (left) and (right), resp.). To confirm that this is not a statistical fluctuation, we ran a higher-resolution experiment with $\tau \in [0.12,0.14]$ using 100 replications. Results are shown in \cref{fig:DetailsRedistribution} (left). Error bars are constructed using \cref{eqn:CI} but with $Z_{1-\alpha/2} = 4.7532$, corresponding to a $p$-value of $10^{-6}$. Thus showing there is a 1 in 1,000,000 chance this is a random statistical fluctuation. 
The symmetry in the asymptotic approximation (and in the binomial distribution itself) suggests that there should be no interaction between tolerance and redistribution, just as in the infinite population case. We hypothesize that these results can be explained in the finite population case as a result of the fact that when \texttt{Reshuffle} is executed, all agents in Group 1 have equal utility, approximately given by $p(0,g_1,g_2)$ (with variance varying inversely with both group size and number of epochs used in the \texttt{MonteCarlo} routine, used to construct a probability of eating). Each agent in Group 2 has a distinct utility. Consequently, this asymmetric relationship is impacting the distribution tails and leading to this behaviour. Interestingly, this behaviour seems to be a purely mesoscopic property. Running the same experiments using a population size of four (i.e., $N = 4$) shows behaviour consistent with expectations. (As before, 100 replications were used.) That is, redistribution does not give either group an advantage. This is shown in \cref{fig:DetailsRedistribution} (right). Constructing an exact mathematical analysis for this behaviour is left to future work.

\begin{figure}[htbp]
\centering
\includegraphics[width=0.45\textwidth]{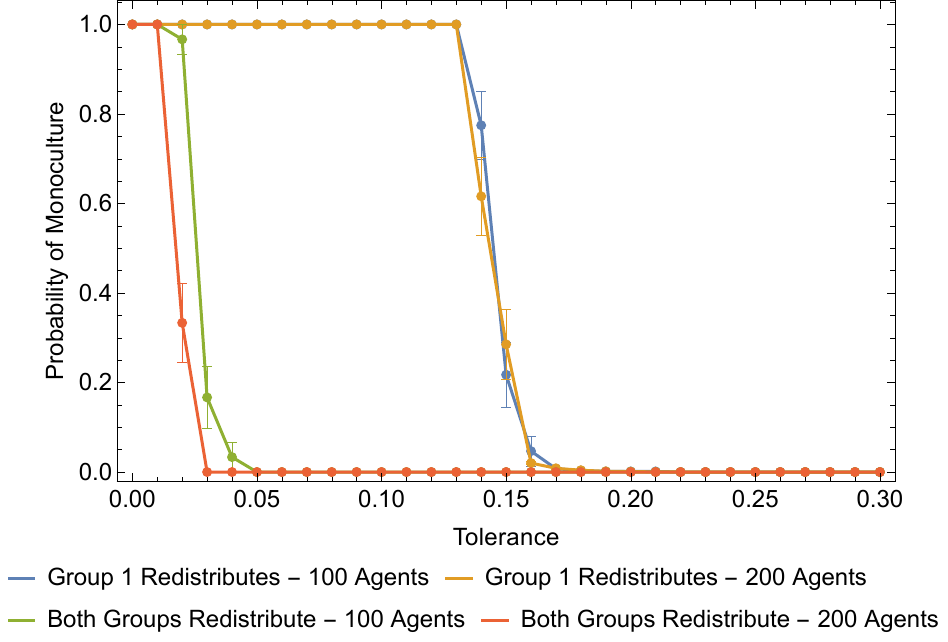} 
\caption{The probability of monoculture formation for different redistribution scenarios shows redistribution causes behaviour more similar to the infinite population case.}
\label{fig:RedistributionMonoculture}
\end{figure}

We note finally that redistribution in multiple groups vs. a single group also affects the probability of monoculture, as shown in \cref{fig:RedistributionMonoculture} (right). 
Comparison with \cref{fig:ProbChangeModel} shows that any redistribution shifts the probability of monoculture formation leftward (as expected), making the system act more like the infinite population case.

\section{Conclusions and Future Directions}\label{sec:Conclusions}
In this paper, we studied a version of the Kolkata Paise Restaurant Problem with multiple dining clubs. Unlike prior work \cite{HBG2023}, we did not reduce the problem to the replicator dynamics, but instead built a model of group exchange where agents exhibited stickiness to their groups. That is, they would tolerate a certain level of inequality before changing groups. We showed that the resulting system exhibits both  regions of stability and that grand coalitions in which all agents join a single dining club are locally asymptotically stable assuming an infinite population. We then explored the impact finite populations had on exchange dynamics, showing that for finite populations, stable regions do not (necessarily) appear and that in the dichotomous case of two dining clubs, all members of the population eventually join one of the two clubs for low enough tolerance to inequality. In this case, when both groups engage in resource (food) redistribution, there is no change to the qualitative nature of the dynamics. However, freeloading by group members from a dining club that does not redistribute can cause the redistributing group to collapse. Finally, when one group redistributes resources and freeloading is prohibited, there is a (surprising) advantage to the redistributing group for certain ranges of iniquity tolerance. Fully explaining the microscopic interactions that lead to this behaviour is left to future work. 

In addition to the future work already noted, there are several future directions worth exploring. Exploring the relationship between this model and the evolutionary game model put forth in \cite{HBG2023} may lead to insights in multi-dining club scenarios. It is known that replicator dynamics are the deterministic limit of stochastic imitation (see e.g., \cite{F24}), however how this plays out in the presence of group (strategy) stickiness as modelled by tolerance to inequality is unclear. Additionally, investigating mesoscopic behaviour when there are $M$ dining clubs when $1 \ll M \ll N$, where $N$ is the number of agents could lead to interesting dynamics, since each monoculture solution (i.e., all agents are in a single club) is an asymptotically attracting solution. Under certain agent update rules, it may be possible to produce chaotic trajectories as each dining club acts like a local magnet in phase space, thus (potentially) mimicking the multi-magnet chaotic pendulum. Additional elements like spatial components or external forces (e.g., fields) may also create tractable but interesting dynamics in both the finite and infinite population cases. 

\section*{Acknowledgements}
A.H and C.G. were supported in part by the National Science Foundation under grant CMMI-1932991. 

\section*{Data and Code Availability}
Mathematica notebooks are provided as supplementary materials and contain the code needed to reproduce the results in this paper as well as an analysis of multi-dining clubs with a replicator dynamic, which does not explicitly appear in this paper.

\bibliographystyle{iopart-num} 
\bibliography{KolkotaPaise}

\providecommand{\newblock}{}
\begin{thebibliography}{10}
\expandafter\ifx\csname url\endcsname\relax
  \def\url#1{{\tt #1}}\fi
\expandafter\ifx\csname urlprefix\endcsname\relax\def\urlprefix{URL }\fi
\providecommand{\eprint}[2][]{\url{#2}}

\bibitem{CMC07}
Chakrabarti B~K, Mitra M and Chakrabarti A~S 2007 {\em arXiv preprint
  arXiv:0711.1639\/}

\bibitem{HBG2023}
Harlalka A, Belmonte A and Griffin C 2023 {\em Physica A: Statistical Mechanics
  and its Applications\/} {\bf 620} 128767

\bibitem{BGCN12}
Biswas S, Ghosh A, Chatterjee A, Naskar T and Chakrabarti B~K 2012 {\em
  Physical Review E\/} {\bf 85} 031104

\bibitem{CCGM17}
Chakrabarti B~K, Chatterjee A, Ghosh A, Mukherjee S, Tamir B {\em et~al.\/}
  2017 {\em Econophysics of the Kolkata Restaurant problem and related games\/}
  (Springer)

\bibitem{CG19}
Chakrabarti A~S and Ghosh D 2019 {\em Journal of Economic Interaction and
  Coordination\/} {\bf 14} 225--245

\bibitem{GC17}
Ghosh D and Chakrabarti A~S 2017 {\em Physica A: Statistical Mechanics and its
  Applications\/} {\bf 483} 16--24

\bibitem{DSC11}
Dhar D, Sasidevan V and Chakrabarti B~K 2011 {\em Physica A: Statistical
  Mechanics and its Applications\/} {\bf 390} 3477--3485

\bibitem{BMM13}
Banerjee P, Mitra M and Mukherjee C 2013 {\em Econophysics of Systemic Risk and
  Network Dynamics\/}  201--216

\bibitem{BM21}
Biswas S and Mandal A~K 2021 {\em Physica A: Statistical Mechanics and its
  Applications\/} {\bf 561} 125271

\bibitem{SC20}
Sinha A and Chakrabarti B~K 2020 {\em Chaos: An Interdisciplinary Journal of
  Nonlinear Science\/} {\bf 30} 083116

\bibitem{GDCM12}
Ghosh A, De~Martino D, Chatterjee A, Marsili M and Chakrabarti B~K 2012 {\em
  Physical Review E\/} {\bf 85} 021116

\bibitem{CCCM15}
Chakraborti A, Challet D, Chatterjee A, Marsili M, Zhang Y~C and Chakrabarti
  B~K 2015 {\em Physics Reports\/} {\bf 552} 1--25

\bibitem{CRS22}
Chakrabarti B~K, Rajak A and Sinha A 2022 {\em Frontiers in Artificial
  Intelligence\/} {\bf 5}

\bibitem{KPA22}
Kastampolidou K, Papalitsas C and Andronikos T 2022 {\em Games\/} {\bf 13} 33

\bibitem{MK17}
Martin L and Karaenke P 2017 The vehicle for hire problem: A generalized
  kolkata paise restaurant problem {\em Workshop on Information Technology and
  Systems\/}

\bibitem{R13}
Ramzan M 2013 {\em Quantum information processing\/} {\bf 12} 577--586

\bibitem{Y10}
Yarlagadda S 2010 Using many-body entanglement for coordinated action in game
  theory problems {\em Econophysics and Economics of Games, Social Choices and
  Quantitative Techniques\/} (Springer) pp 44--51

\bibitem{GCCC14}
Ghosh A, Chatterjee A, Chakrabarti A~S and Chakrabarti B~K 2014 {\em Physical
  Review E\/} {\bf 90} 042815

\bibitem{M12}
Morris P 2012 {\em Introduction to game theory\/} (Springer Science \& Business
  Media)

\bibitem{CZ98}
Challet D and Zhang Y~C 1998 {\em Physica A: Statistical Mechanics and its
  applications\/} {\bf 256} 514--532

\bibitem{HZDH12}
Huang Z~G, Zhang J~Q, Dong J~Q, Huang L and Lai Y~C 2012 {\em Scientific
  reports\/} {\bf 2} 703

\bibitem{arthur1994}
Arthur W~B 1994 {\em The American Economic Review\/} {\bf 84} 406--411

\bibitem{decara1999}
de~Cara M, Pla O and Guinea F 1999 {\em The European Physics Journal B\/} {\bf
  10} 187--191

\bibitem{FGH02}
Farago J, Greenwald A and Hall K 2002 Fair and efficient solutions to the
  {S}anta {F}e bar problem {\em Proceedings of the Grace Hopper Celebration of
  Women in Computing\/}

\bibitem{challet2004}
Challet D, Marsili M and Ottino G 2004 {\em Physica A\/} {\bf 332} 469--482

\bibitem{AFGJ13}
Antoniadis P, Fdida S, Griffin C, Jin Y and Kesidis G 2013 Distributed medium
  access control with dynamic altruism {\em Ad Hoc Networks: 4th International
  ICST Conference, ADHOCNETS 2012, Paris, France, October 16-17, 2012, Revised
  Selected Papers 4\/} (Springer) pp 29--42

\bibitem{AFGJ13a}
Antoniadis P, Fdida S, Griffin C, Jin Y and Kesidis G 2013 {\em EURASIP Journal
  on Wireless Communications and Networking\/} {\bf 2013} 1--12

\bibitem{GK14}
Griffin C and Kesidis G 2014 Behavior in a shared resource game with
  cooperative, greedy, and vigilante players {\em 2014 48th Annual Conference
  on Information Sciences and Systems (CISS)\/} (IEEE) pp 1--6

\bibitem{GCMC10}
Ghosh A, Chatterjee A, Mitra M and Chakrabarti B~K 2010 {\em New Journal of
  Physics\/} {\bf 12} 075033

\bibitem{BSC24}
Biswas A, Sinha A and Chakrabarti B~K 2024 {\em Indian Journal of Physics\/}
  1--7

\bibitem{GSC10}
Ghosh A, Sundar~Chakrabarti A and Chakrabarti B~K 2010 Kolkata paise restaurant
  problem in some uniform learning strategy limits {\em Econophysics and
  Economics of Games, Social Choices and Quantitative Techniques\/} (Springer)
  pp 3--9

\bibitem{H00}
Hethcote H~W 2000 {\em SIAM review\/} {\bf 42} 599--653

\bibitem{EGB16}
Ermentrout G~B, Griffin C and Belmonte A 2016 {\em Physical Review E\/} {\bf
  93} 032138

\bibitem{F24}
Fontanari J~F 2024 {\em Europhysics Letters\/} {\bf 146} 47001

\end{thebibliography}
\end{document}